\documentclass[journal]{IEEEtran}
\usepackage{subfig}
\usepackage{cite}
\usepackage{amsmath,amssymb,amsfonts}
\usepackage{graphicx}
\usepackage{textcomp}
\usepackage{xcolor}
\def\BibTeX{{\rm B\kern-.05em{\sc i\kern-.025em b}\kern-.08em
    T\kern-.1667em\lower.7ex\hbox{E}\kern-.125emX}}
\usepackage{booktabs}
\usepackage{algorithm}
\usepackage{algpseudocode}
\usepackage{pgfplots}
\usepackage{tabularx}
\usepackage{cite}
\usepackage[top=1.0in, bottom=1.1in, inner=0.8in, outer=0.8in]{geometry}
\usepackage{multirow}
\begin{document}
%
% paper title
% Titles are generally capitalized except for words such as a, an, and, as,
% at, but, by, for, in, nor, of, on, or, the, to and up, which are usually
% not capitalized unless they are the first or last word of the title.
% Linebreaks \\ can be used within to get better formatting as desired.
% Do not put math or special symbols in the title.
\title{Entropy-Based Dynamic Programming \\
for Efficient Vehicle Parking}
%
%
% author names and IEEE memberships
% note positions of commas and nonbreaking spaces ( ~ ) LaTeX will not break
% a structure at a ~ so this keeps an author's name from being broken across
% two lines.
% use \thanks{} to gain access to the first footnote area
% a separate \thanks must be used for each paragraph as LaTeX2e's \thanks
% was not built to handle multiple paragraphs
%

\author{Jean-Luc Lupien $^{1}$\textsuperscript{\textdaggerdbl}, Abdullah Alhadlaq $^{1}$\textsuperscript{\S}, Yuhan Tang $^{1}$\textsuperscript{\textdagger}*,  Yan Wu$^{1}$, Jiayu Joyce Chen$^{1}$, \\  Yutan Long$^{1}$

\thanks{$^{*}$Corresponding author. \textsuperscript{\textdagger \textdaggerdbl \S} These authors contributed equally.}
\thanks{$^{1}$Department of Civil and Environmental Engineering, University of California, Berkeley,
        Berkeley, CA 94720, USA
        {\tt\small  jllupien@berkeley.edu, alhadlaq@berkeley.edu, yhtang@berkeley.edu, yan\_wu@berkeley.edu, chenjiayu@berkeley.edu, ytlong@berkeley.edu
        }
        }

}

\maketitle

% As a general rule, do not put math, special symbols or citations
% in the abstract or keywords.
\begin{abstract}
% With limited information for the driver and the huge demand of parking, drivers have been wasting lots of time wandering inside garages to find spots. We introduces an innovative approach to optimizing vehicle parking through an entropy-based dynamic programming model. In urban environments, where parking is a significant source of congestion and inefficiency, our methodology offers a structured solution to minimize the time spent by drivers in finding parking spaces. Utilizing entropy models derived from statistical mechanics, we predict the distribution of available parking spots across different levels of a multi-story parking garage based on the information observed. Our Temperature-Informed Parking Policy (TIPP) allows us to dynamically adjust parking strategies based on real-time data and historical patterns. We further develop a dynamic programming framework that guides vehicles to the optimal floor, significantly reducing the overall time spent searching for a spot. TIPP is compared with different  conventional parking strategies in a simulated environment, showcasing superior performance in efficiency. We also compared the result of TIPP with the optimal strategies with full information, which is the best possible solution, indicating our well performance only based on partial data. The results highlight the potential of integrating TIPP in real-world applications, paving the way for smarter, more efficient urban landscapes.

In urban environments, parking has proven to be a significant source of congestion and inefficiency. In this study, we propose a methodology that offers a systematic solution to minimize the time spent by drivers in finding parking spaces. Drawing inspiration from statistical mechanics, we utilize an entropy model to predict the distribution of available parking spots across different levels of a multi-story parking garage, encoded by a single parameter: temperature. Building on this model, we develop a dynamic programming framework that guides vehicles to the optimal floor based on the predicted occupancy distribution. This approach culminates in our Temperature-Informed Parking Policy (TIPP), which not only predicts parking spot availability but also dynamically adjusts parking assignments in real-time to optimize vehicle placement and reduce search times. We compare TIPP with simpler policies and the theoretical optimal solution to demonstrate its effectiveness and gauge how closely it approaches the ideal parking strategy. The results highlight the potential of integrating TIPP in real-world applications, paving the way for smarter, more efficient urban landscapes.

\end{abstract}

% Note that keywords are not normally used for peerreview papers.
% \begin{IEEEkeywords}
% Bus Operation, Reinforcement Learning, Policy Proximal Optimization, Curriculum Learning
% \end{IEEEkeywords}

% For peer review papers, you can put extra information on the cover
% page as needed:
% \ifCLASSOPTIONpeerreview
% \begin{center} \bfseries EDICS Category: 3-BBND \end{center}
% \fi
%
% For peer review papers, this IEEEtran command inserts a page break and
% creates the second title. It will be ignored for other modes.
\IEEEpeerreviewmaketitle

% The very first letter is a 2 line initial drop letter followed
% by the rest of the first word in caps.
% 
% form to use if the first word consists of a single letter:
% \IEEEPARstart{A}{demo} file is ....
% 
% form to use if you need the single drop letter followed by
% normal text (unknown if ever used by the IEEE):
% \IEEEPARstart{A}{}demo file is ....
% 
% Some journals put the first two words in caps:
% \IEEEPARstart{T}{his demo} file is ....
% 
% Here we have the typical use of a "T" for an initial drop letter
% and "HIS" in caps to complete the first word.
\section{Introduction}
 \IEEEPARstart{I}{n} modern urban landscapes, parking infrastructure plays a crucial role in addressing the burgeoning challenges of congestion, pollution, and land utilization. As cities continue to expand, the demand for effective parking solutions increases. Notably, Americans spend an average of \$72.7 billion searching for parking spots \cite{inrix}. Consequently, optimizing the management of both urban parking lots and internal parking lot structures is imperative. While urban parking lots cater to on-street and surface parking within city environments, internal parking structures, often multi-level facilities, require distinct strategies due to their vertical design and capacity constraints. Differentiating these contexts allows for tailored optimization approaches, enhancing overall efficiency and user experience in diverse urban settings.

Traditional approaches to parking structure design and optimization often prioritize maximizing capacity within limited building areas, thereby neglecting broader implications such as traffic flow optimization, environmental sustainability, and integration with emerging vehicle capabilities. 
There are many problems associated with the difficulty with parking. It is estimated that 63 million Vehicle Miles Traveled (VMT) are generated every year in only the city of Chicago because of inefficient parking strategies \cite{parking_games}. 
The current literature addresses many aspects of the parking problem. The first aspect is in parking spot assignments, with many focused on smart parking systems that guide drivers to the nearest available space. Abidi et al proposed an algorithm to find the best parking spot within a region. They proposed a hybrid genetic assignment search procedure (HGASP) with a time restriction, which could reduce up to 500 thousand kilometers traveled by vehicles within a city for all vehicles \cite{abidi2015}. Chen et al used a prediction scheme of parking lot availability based on sample parking spaces. They used Fuzzy logic to predict available parking spaces \cite{chen2013}. On a smaller scale, Yan et al calculated the optimal path to the parking spot using the Dijkstra algorithm for each vehicle, and the information on the assignment and path was sent to the drivers’ cell phones. Their experimental result showed a 5-minute time reduction in cruising to find parking \cite{yan2017}.  Ayala et al. use a Gravity-based Parking Algorithm with a graph representation of the cars in a parking lot with both complete and partial information. They found the best case to be able to reduce 25\% of vehicle cruising time \cite{parking_games}.

Recent advancements in artificial intelligence, specifically reinforcement learning (RL) have shown promise in solving complex optimization problems in dynamic and uncertain environments. Several studies have applied RL to optimize parking space allocation and reduce search time. Zhang et al. proposed a multi-agent deep reinforcement learning framework for online parking assignments with both connected and non-connected vehicles with partial observations of parking demand. Their results showed that, if user parking information of connected vehicles was provided, their algorithm could reduce 15\% of the total time that users spend to find parking and travel to their destination as compared with baselines \cite{Zhang2022}. Khalid, et al. created a situation where an Autonomous Vehicle is used to pick up and drop off users at their required spots in the car park, and drive around the city autonomously. They used a Double-Layer Ant Colony Optimization algorithm and Deep Q-learning Network to model a dynamic environment \cite{Khalid2022}. Their result shows considerable performance as compared to random route finding. 
However, these genres of literature often require hypothetical capabilities from vehicles, such as connectivity and full information sharing, which are infeasible today. They also often do not account for the unique requirements of mixed-use buildings, where the temporal distribution of parking demand varies significantly between different user groups. On the other hand, entropy model is used in transportation prediction, based on information entropy theory, Hu et al. presents a model to evaluate the difficulty of failure disposal \cite{hu2019research}.

% \paragraph{Contributions}
Our contributions are three-fold. Firstly, we present a single-parameter model for predicting the distribution of parking spots in a parking structure. Specifically, given the \emph{temperature} of a parking lot, our model can be used to predict occupancy within the parking lot. Secondly, we propose a dynamic programming approach to determining the best parking strategy for a car entering the parking structure given a distribution of spots. Finally, we illustrate the performance of the combination of the single-parameter prediction model and the dynamic programming approach on a simulated parking lot. The presented framework is shown to outperform common human parking strategies.

\section{Problem Formulation}
For simplicity, we have selected an N-layer parking garage as our research subject, with each level having its own distinct probability of containing an available parking spot, denoted as $p_i$.

The reason for selecting this type of parking garage is rooted in real-life scenarios where vehicles are restricted to moving either upward or downward between adjacent floors. For this research, we assume that the vehicles can only descend to the next floor, while passengers can only traverse to the upper level on foot. We denote $t_2$ as the time taken by a passenger to ascend one floor on foot and $t_3$ as the time required for a vehicle to descend one floor separately. Additionally, we employed $t_1$ to represent the duration spent by a car scanning a floor for an available spot.

\begin{figure}
    \centering
    \includegraphics[width=0.35\textwidth]{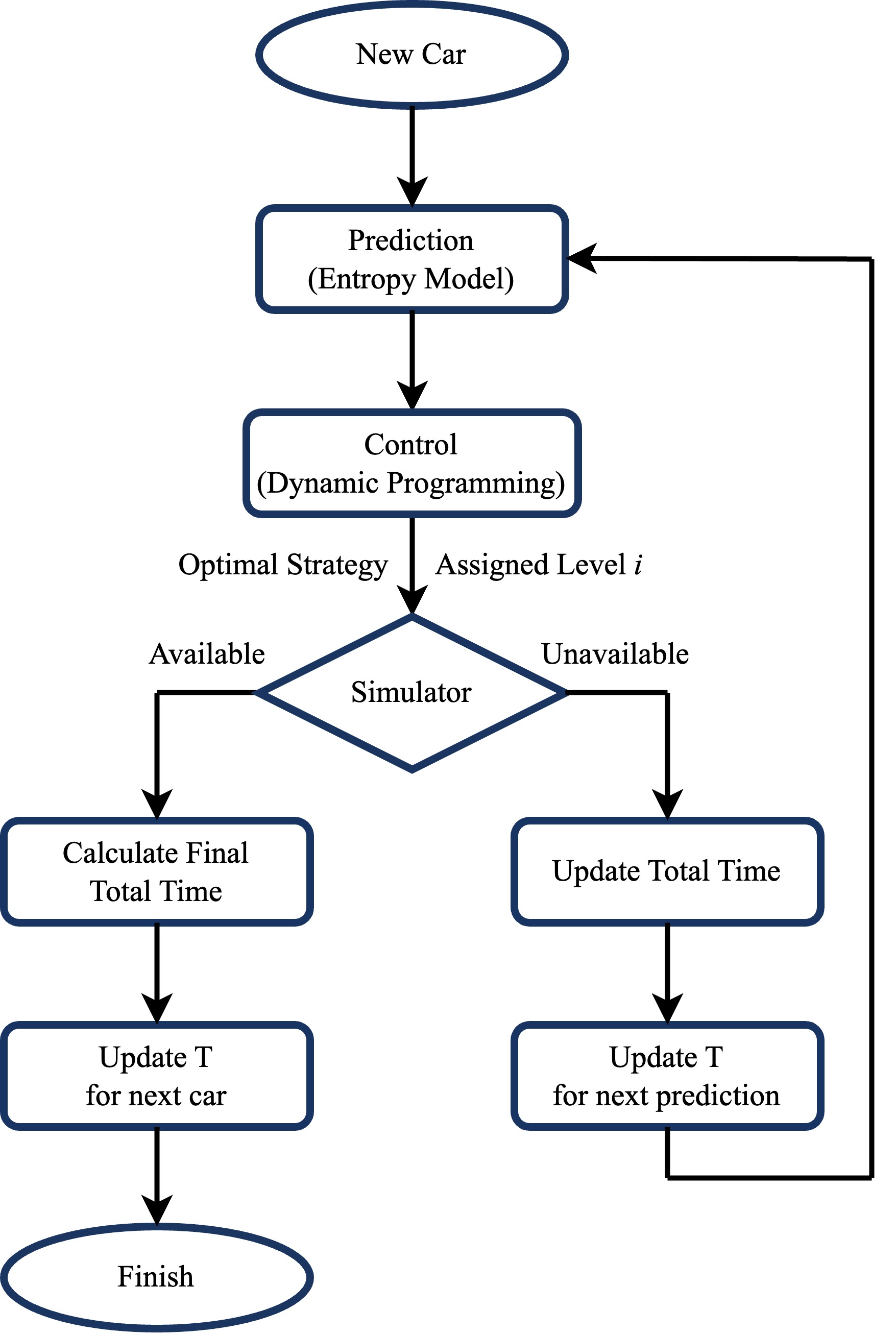} 
    \caption{The framework of this research}
    \label{fig:framework}
\end{figure}

To address this complex problem, our project contains three essential components: prediction, control, and simulation. The framework of this research is shown in Figure \ref{fig:framework}.

In the prediction part, we try to predict the distribution of parking occupancy across different levels. To achieve this, we employ the entropy model, which analyzes historical data and real-time inputs to predict the likelihood of finding available parking spaces on each floor.

In the control part, our goal is to generate an optimal strategy for assigning vehicles to specific levels within the parking garage. Here, we leverage the power of dynamic programming, to generate strategies that maximize efficiency and minimize the time required for parking and accessing vehicles.

In the simulator, to explore the dynamics of parking behavior and assess the efficiency of various strategies, we developed a parking simulator using Python. In this section, we introduce the structure and functionality of our simulation model, encapsulated within the \texttt{ParkingGarage} class. It adopts an object-oriented design with modules for vehicle management, parking spot management, and parking lot status updates. The \texttt{ParkingGarage} class serves as the core of the simulation, initializing parameters including the number of levels and parking spots per level, and providing features like parking car simulation, dynamic renewal to mimic varying flow scenarios, and energy consumption calculation for parked cars. Additionally, a visualization method generates graphical representations of the parking garage's occupancy status using Matplotlib, aiding in intuitive analysis.

\section{Prediction - Entropy Model} \label{entropy}

To model the distribution of cars in parking lots, we turn to statistical mechanics for inspiration. Specifically, we posit that the distribution of resembles that predicted by the canonical ensemble. The canonical ensemble describes an open system placed in thermal equilibrium with a heat bath \cite{statmech}. At thermal equilibrium, the system will have a specific occupancy distribution of its states entirely determined by the energy associated with each state. Low-energy states will have a high probability of occupancy as compared to high-energy states. At low temperatures, states will have a low probability of being occupied. To adapt the a parking lot to this framework, we make two assumptions:

\begin{enumerate}
    \item The parking lot is in \emph{thermal equilibrium} with a heat bath. In this case, this means that the the average inflow and outflow of cars from the parking lot are equal.
    \item The desirability of each parking spot can be described by a scalar quantity analogous to \emph{energy}.
\end{enumerate}

Under these assumptions, the canonical ensemble state distribution can be employed where the probability of a state $i$ being occupied is given by the following formula:
\begin{equation} \label{eq:entropy}
p(i) = \frac{2e^{\frac{E(i)}{k_BT}}}{1+e^{\frac{E(i)}{k_BT}}},\end{equation}
where $E(i)$ is the energy of state $i$, $T$ is the temperature of the system, and $k_B$ is Boltzmann's constant. We remark that the only independent parameter is temperature.

Using this framework, we can associate an energy level with each parking spot in the parking lot based on its attractiveness. A lower energy represents a more desirable spot. We suppose that the energy of a particular parking spot is linked only to its distance to a point of interest (entrance, elevator, exit, etc). Empirically, we have found that supposing an energy proportional to the square of the distance yields the best fit with respect to the data. The temperature will then represent how busy the parking lot is: higher temperatures indicate higher utilization and therefore a higher fill rate. Given a temperature and the energy of any spot in the parking lot will then directly yield the probability of occupancy of this spot. 

To validate this model, 10 sample parking lots of different sizes and occupancy were considered from Antioch and Dublin, CA, Moab, UT, Colorado Springs, CO, and Louisville, KY. We ensure there was only one point of interest for the entire parking lot, for example, the single entrance to a shopping center. The distance between every parking spot and the point of interest was then calculated. To account for the difference in the sizes of parking lots, these distances were then normalized for each parking lot. The energy of that parking spot was then associated with the square of the normalized distance. The temperature of the parking lot was then determined by minimizing the mean squared error (MSE) of the Entropy Model using gradient descent. 

\begin{figure}
    \centering
    \includegraphics[width=0.35\textwidth]{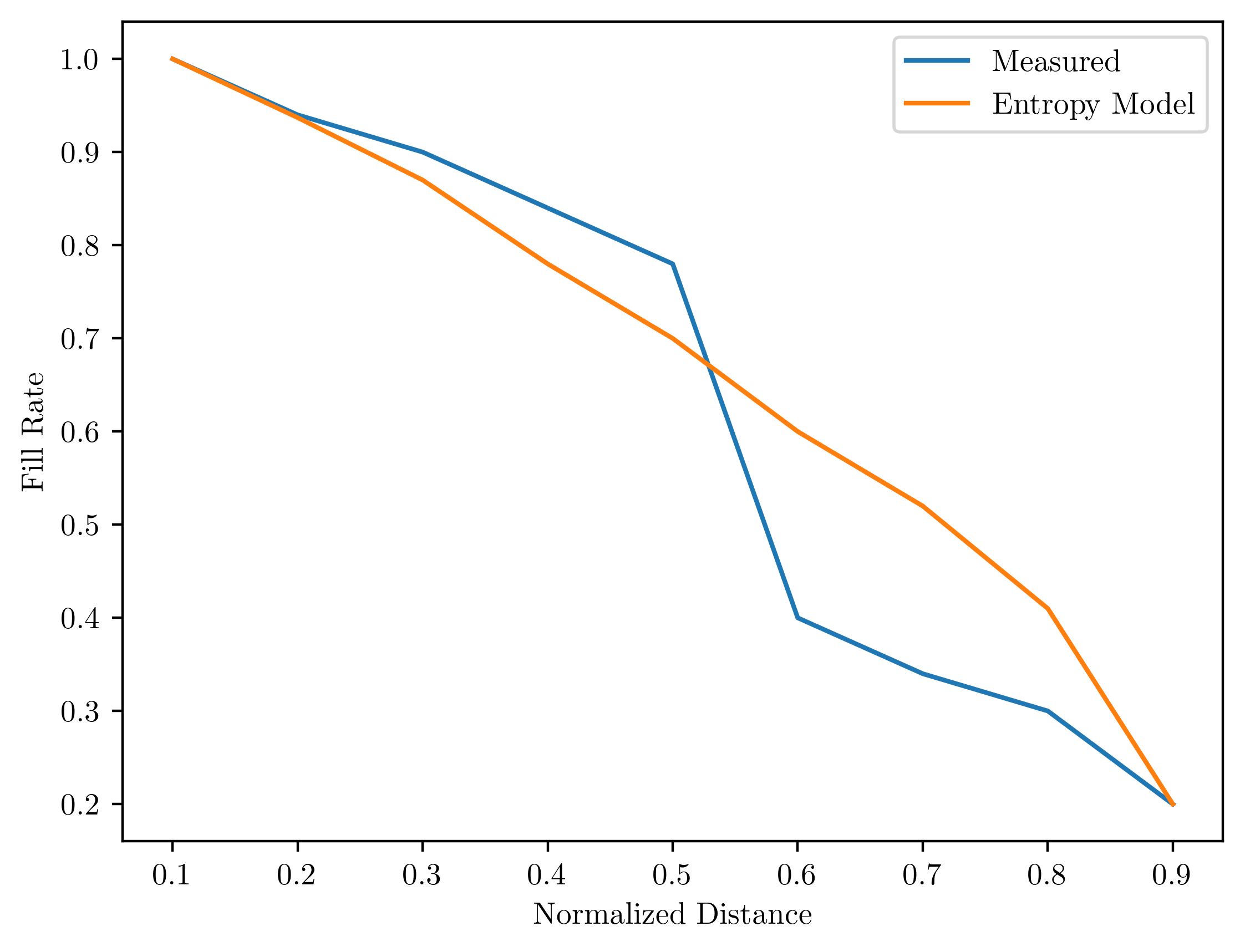} 
    \caption{Predicted and Measured Fill Rate as a Function of Distance}
    \label{fig:fit}
\end{figure}

The aggregate results for all 10 parking lots are presented in Figure \ref{fig:fit}. Every parking lot's temperature was determined individually, then results were pooled. We observe that the Entropy Model has an excellent fit to the data. Indeed, the calculated MSE over the entire dataset is $0.173$. This suggests that the Entropy Model is an accurate representation of the distribution of parking spaces.

\begin{figure}
    \centering
    \includegraphics[width=0.35\textwidth]{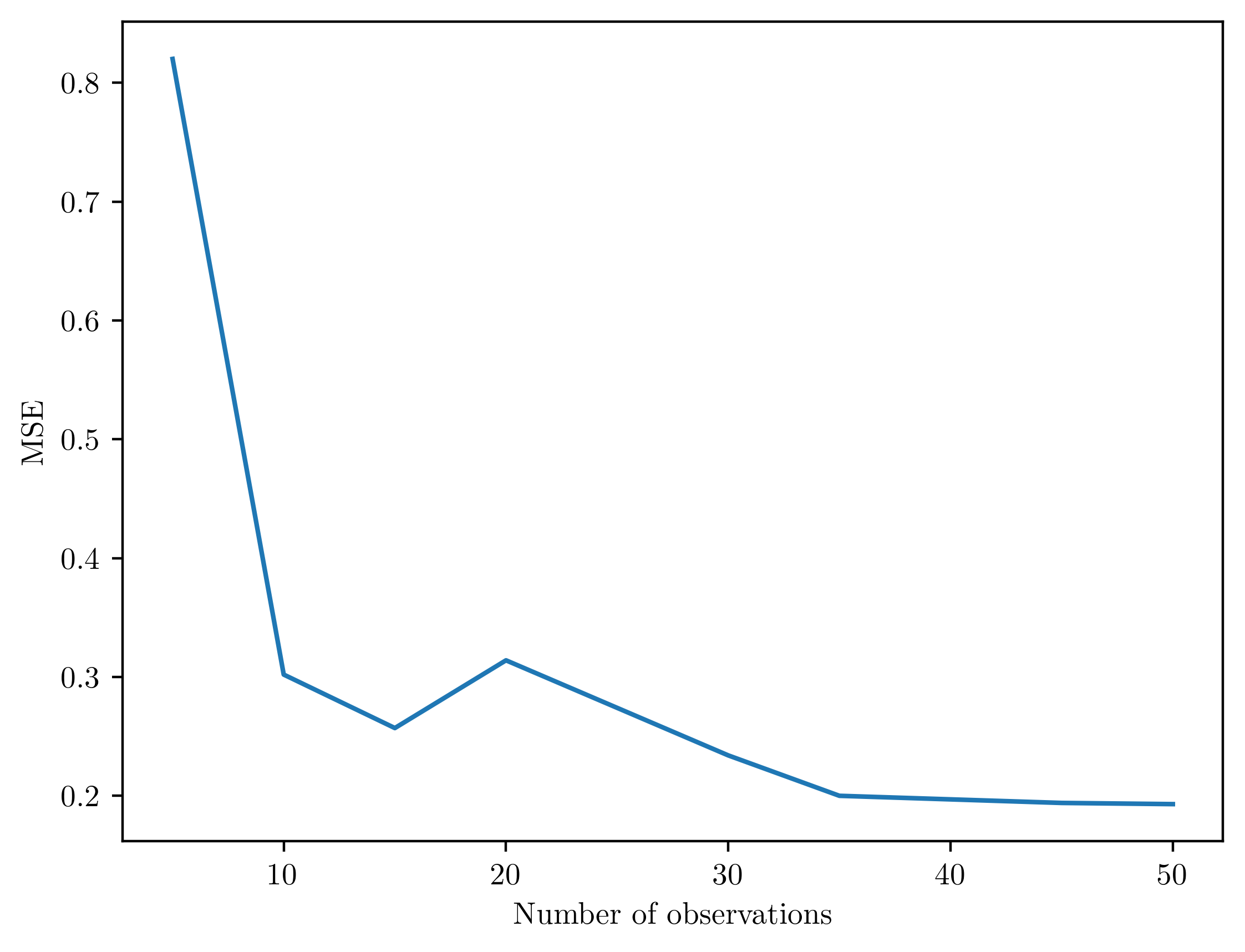} 
    \caption{MSE as a Function of Observations in Sample}
    \label{fig:MSE}
\end{figure}

A major advantage of using this single-parameter model lies in its sample efficiency. We now show that very few observations are necessary to obtain an accurate representation of the entire parking lot. To do this, we calculate the best-fit temperature for a sample of parking spots. From this temperature, we then calculate the mean squared error on the entire parking lot from the Entropy Model. The results of this experiment are presented in Figure \ref{fig:MSE}. The X-axis represents the sample size for temperature estimation out of a total of 105 parking spots. We observe that even from 10 observations (less than 10\% of total spots), the Entropy Model produces a high-quality approximation of the entire parking lot.

The high accuracy of the model combined with its sample efficiency justifies its selection as a predictive model to aid dynamic decision-making. From a few observations, a realistic temperature update can be calculated from which the entire distribution can be approximated. This will be useful for parking decisions which are presented in the next section. Furthermore, this model has advantages over full-information parking systems where, for example, the total number of vehicles in the parking structure is known. The sample efficiency of the model means that a lot of costs can be saved when compared to such infrastructure-heavy installations. Because the information is conveyed uniquely by the temperature, this also avoids any privacy concerns over information sharing.

\section{Control - Dynamic Programming}

We formulate our problem by considering a parking garage with $N$ floors, where the probability of finding a parking spot on the $i$-th floor is given by $p_i$.

A car starts its position at the $0$-th floor and then decides which floor to go to. After going to floor $i$, the car spends $t_1$ time scanning the floor for a spot. If it finds a spot, then it parks there, and the passenger will spend $it_2$ time to go back to the entrance. If however, no spot was found on floor $i$, then the car decides to go to another floor, say $j$, further down. This means that the car will spend $(j-i)t_3$ time to go down to that floor. The process continues until a car finds a spot. Please notice that although the back time of
passengers ($it_2$) won't influence the parking time for the next vehicle, it reflects the closeness of this parking spot to the exit, which will definitely affect parking behaviour. Therefore, it's taken into consideration.

The goal is to construct a control action $u(i)$ which dictates the best floor to go to from floor $i$ minimizing the expected total time spent finding a spot. Let's assume that a car makes a series of $n$ actions that end with the car finding a parking spot at floor $a_n$ which is the final action. Then, the total time $T$ it takes is described in the equation below.
\begin{equation} \label{eq:tota_time}
    T = nt_1 + a_n (t_2 + t_3)
\end{equation}
Using this definition, we want to minimize $E[T]$. This can therefore be considered a Markov decision process where the state space $S$ is the set of current floor $i$ and temperature $T$, so $S=\{i, T \mid i \in \{1, 2, ..., N\}\}$. The action space at floor $i$ is the set of floors beneath $i$, namely $\{i+1, i+2, ..., N\}$. We are assuming in this description that the driver can only go to floors beneath the ones they are at currently, which is a reasonable assumption to make especially if $t_3 << t_1$ (it takes more time to scan a floor looking for a spot, than it does traversing up one floor on foot, which is usually done through an elevator in seconds). This approximation allows us to solve the problem optimally using dynamic programming.

%We form the parking problem as a dynamic programming problem. Our case is on a multi-layer parking lot. Let the car entered in floor 0. Each layer is denoted as 0, 1, 2, 3, ..., $N-1$, $N$. Assume we already get the probability of finding a parking space at level $i$, called $p_i$.% 

Starting from the bottom floor (boundary condition), we can calculate the total parking time to be:
\begin{equation} \label{value}
f(N)=t_1+Nt_2
\end{equation}
If we are instead on the floor right above N, which is floor $N-1$, the total parking time would be:

\begin{equation} \label{value}
\begin{split}
f(N-1) &= p_{N-1}\left(t_1+(N-1) t_2\right) + \\
       &\quad \left(1-p_{N-1}\right)\left(t_1+t_3+f(N)\right)
\end{split}
\end{equation}

Generalizing this for any floor, let our objective function $f(i)$ be the minimum expected total time starting from floor $i$. It's equal to the time it takes if there's a parking spot on floor $i$ (case 1) plus the time it takes if there's no parking spot on floor $i$ (case 2), and then times the cases of their probabilities. If it's in case 1, the time it takes would be the driving time  $t_1$ and walking time $t_2$ times the floor that passed by, which is $i$ here. If it's in case 2, the car will spend time $t_1$ on that floor and then explore a lower floor $j$. The driver then takes the time to drive to floor $j$ and spends another time $f(j)$ to find a spot. The minimum time it takes to find a spot on another floor is added to the time spent in case 2. The equation of $f(i)$ is formed as:

% \begin{equation} \label{value}
% f(i)=p_i\left(t_1+i t_2\right)+\left(1-p_i\right) (t_1+\min _{j>i}\left((j-i) t_3+f{(j)}\right))
% \end{equation}

\begin{equation} \label{value}
\begin{split}
f(i) &=p_i\left(t_1+i t_2\right)+ \\& 
\left(1-p_i\right) (t_1+\min _{j>i}\left((j-i) t_3+f{(j)}\right))
\end{split}
\end{equation}

Let the strategy be a number indicating which floor the car goes to. For the car that starts at floor $i$, we want it to go to the floor that is closest to the entrance, i.e. the floor with the lowest $f(i)$ value. Hence, we can construct the optimal control action as:
\begin{equation} \label{policy}
u(i) = \underset{j>i}{\operatorname{argmin}}\left[(j-i) t_3+f(j)\right]
\end{equation}

At each floor visit, we update the values of the probabilities $p_i$ using the method described in section~\ref{entropy}, run the dynamic program described in equations \ref{value} and \ref{policy}, and obtain the best strategy for parking. 

% An interesting extension to this is removing our assumption that users will only visit floors underneath, and might decide to go back to previous floors (if, for example, they discover that the parking lot is very empty). then, the problem becomes more complex because we have cycles in the optimization step, and we need to do iterative dynamic programming to solve it. We can perform value iteration on the equation below:
% \begin{equation}
% v(i)=p_i\left(t_1+i t_2\right)+\left(1-p_i\right) (t_1+\min _{j}\left(|j-i| t_3+v{(j)}\right))
% \end{equation}
% This would require multiple iterations until convergence, and then similarly to what we did before, we can construct the optimal policy as shown in the equation below:
% \begin{equation}
% \pi(i) = \underset{j}{\operatorname{argmin}}\left[|j-i| t_3+v(j)\right]
% \end{equation}
% In our current experiments, we plan to compare the results of both approaches. We expect that the assumption holds given $t_1 >> t_3$.

\section{Experiments}

In our experiments, we compare multiple parking strategies applied within a 10-level garage, each level containing 30 parking spots. The first strategy, referred to as the 'Benchmark Policy,' involves a straightforward approach where the vehicle starts at the level nearest to the entrance (top floor) and moves sequentially downwards through each level. The car scans each floor for an available parking spot and parks at the first available spot it encounters. This policy values closeness to the entrance the most.

The second strategy, named the 'Inverse Policy,' mirrors the Benchmark Policy but in reverse order. Here, the vehicle starts from the bottom floor and progresses upwards towards the entrance. Similar to the Benchmark Policy, the car scans each floor for available spots and parks at the first spot it finds, effectively trying to avoid scanning floors by seeking likely empty ones. The third strategy is our proposed method Temperature-Informed Parking Policy (TIPP), which utilizes the dynamic programming framework described earlier.

\begin{figure}
    \centering
    \includegraphics[width=0.4\textwidth]{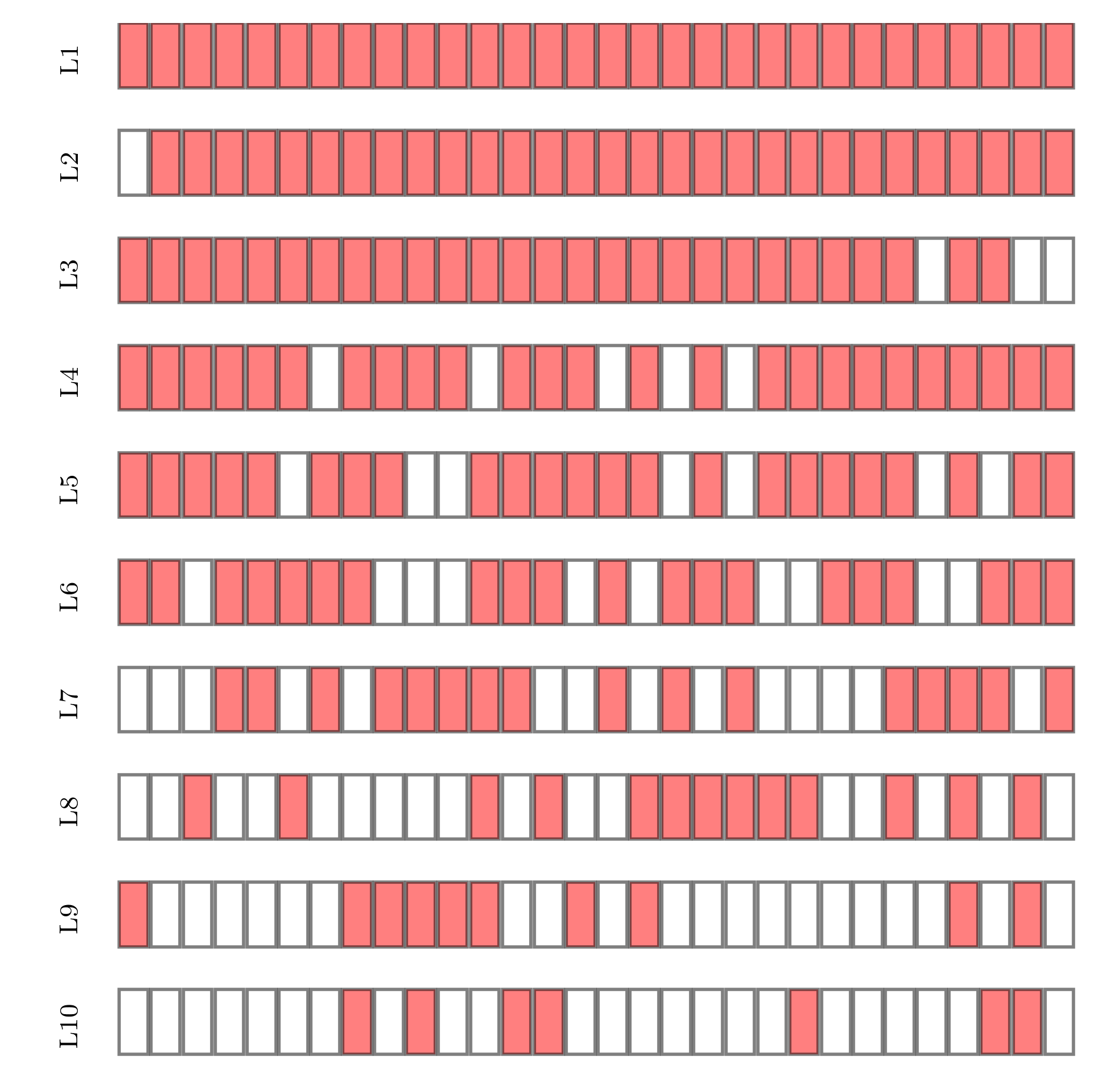}
    \caption{Depiction of the parking garage utilized in the experimentation phase. This is a 10-level parking garage, each level featuring 30 parking spots. Red spots indicate occupied spaces, and white spots represent empty ones. The occupancy distribution is determined using a temperature of 0.5 to simulate the fill rate as per the Entropy Model}
    \label{fig:exp_parking_lot}
\end{figure}

All strategies are evaluated against the 'Optimal Policy', which operates with full visibility of every available parking spot, enabling it to allocate the car to the most advantageous spot immediately. We simulate a parking lot environment at a fixed temperature of 0.5, which determines the occupancy of each level based on the probabilities derived from Eq.~\ref{eq:entropy}. These probabilities dictate the fill rates for each level, subsequently adjusted to the nearest whole number. The energy function for each floor is defined as \(E(i) = (i/N)^2\), where \(i\) ranges from 1, representing the floor closest to the entrance, to 10, the floor farthest away, and $N=10$ for normalization. A depiction of the distribution of the occupancy for the garage in question is shown in Figure~\ref{fig:exp_parking_lot}.

\begin{figure}
    \centering
    \includegraphics[width=0.4\textwidth]{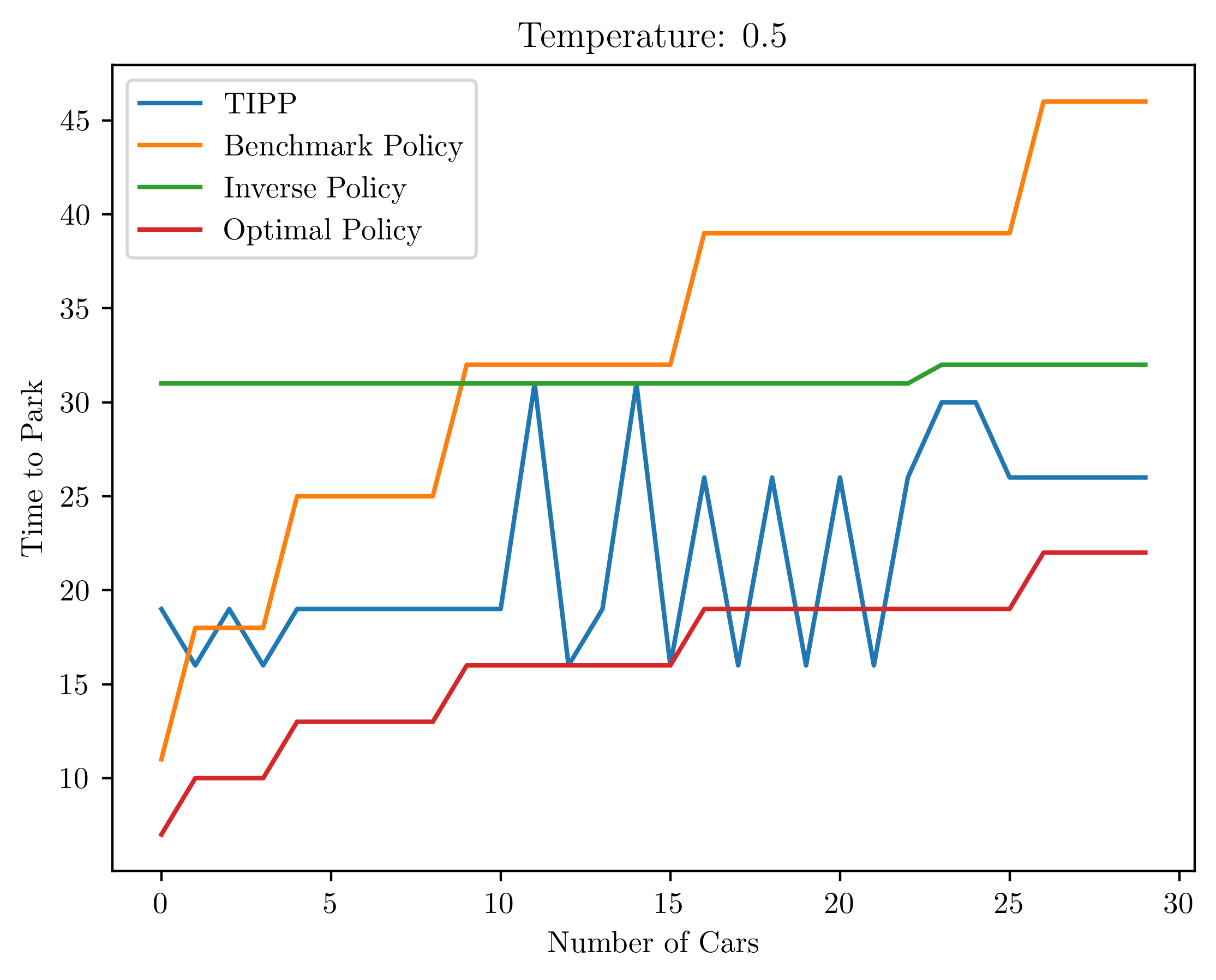}
    \caption{Comparison of the time taken to park for 30 sequential cars under four different parking policies in a garage simulated at a temperature of 0.5.}
    \label{fig:per_car}
\end{figure}

To compare the parking policies, we sequentially insert 30 cars into the garage using one policy, then reset the garage to its initial state (with a temperature of 0.5) and repeat the insertion process for each policy. Figure~\ref{fig:per_car} illustrates the time it takes for each of the cars to find a parking spot and for the driver to return to the entrance, as calculated based on Eq.~\ref{eq:tota_time}.

As illustrated in Figure~\ref{fig:per_car}, the time to park under the Benchmark Policy shows a monotonically increasing trend. This increase occurs because each new car must scan the already full floors before moving to other floors with available slots. Conversely, the Inverse Policy's time remains constant for an extended period, as it begins by filling the lowest floor and progresses upward. On the other hand, TIPP exhibits a more dynamic behavior, consistently outperforming the other two policies, except at the outset where the Benchmark Policy initially benefits from the availability of a few empty spots on the top level.

An intriguing observation is the noticeable fluctuations in parking times under TIPP. Despite the deterministic output of the policy, these variations resemble a balance of exploration and exploitation behaviors. As the chosen optimal floor approaches full capacity, the observed temperature becomes higher than the actual temperature. This discrepancy leads the policy to recommend a more distant but emptier floor, subsequently allowing for a correction in the temperature reading and enabling the exploitation of strategically better floors for parking. Furthermore, although there are instances where the parking time of TIPP is temporarily less than the optimal policy, it is important to note that this does not indicate superiority over the Optimal Policy. Instead, it reflects the timing of when the optimal spots were utilized, with the Optimal Policy having accessed these premium positions earlier in the process.

\begin{figure*}[!t]
\centering
\subfloat[Parking garage under light occupancy]{%
  \includegraphics[width=0.32\textwidth]{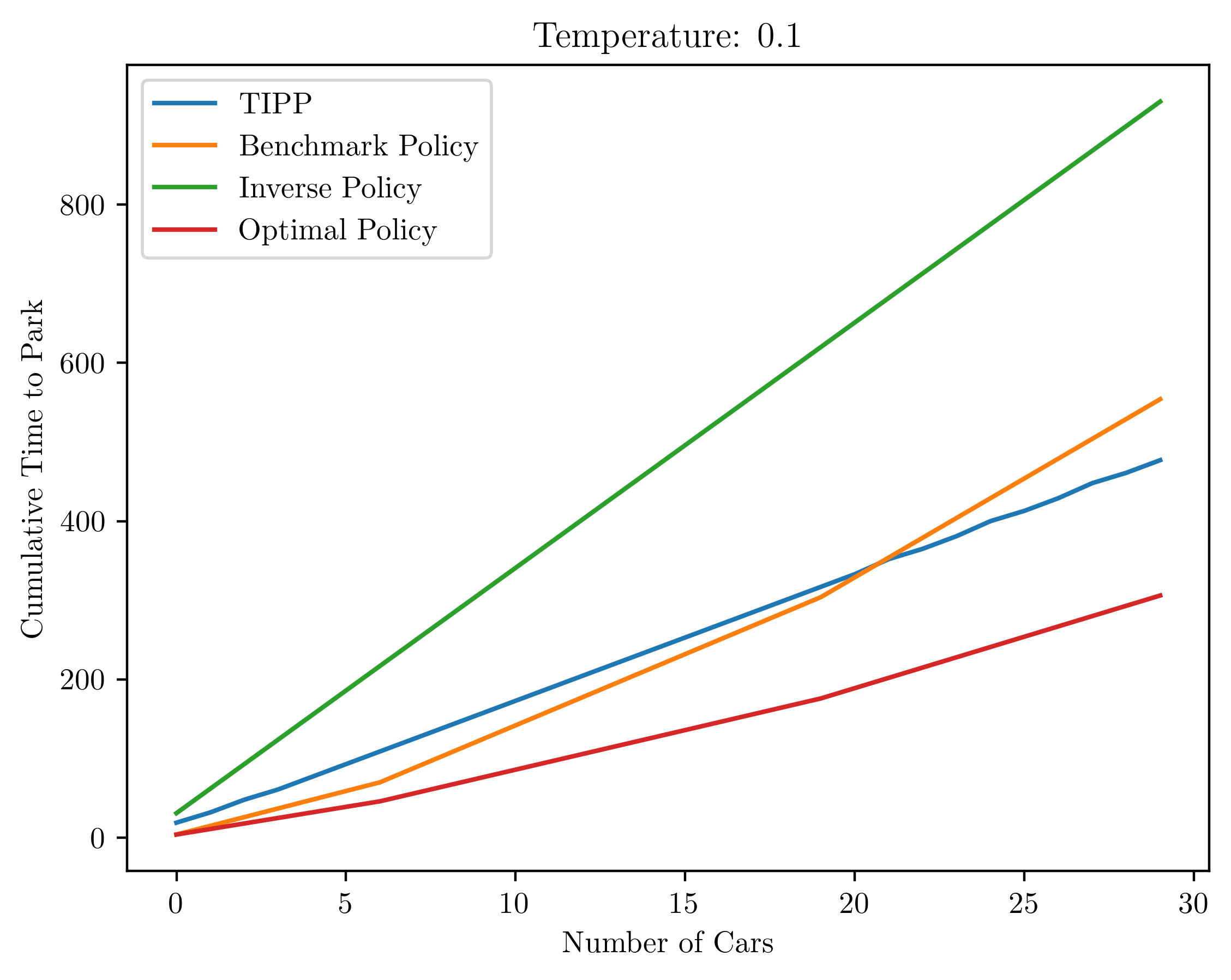}
  \label{fig:cumulative_a}
}
\hfill  % This ensures that the figures spread out to fill the line
\subfloat[Parking garage under medium occupancy]{%
  \includegraphics[width=0.32\textwidth]{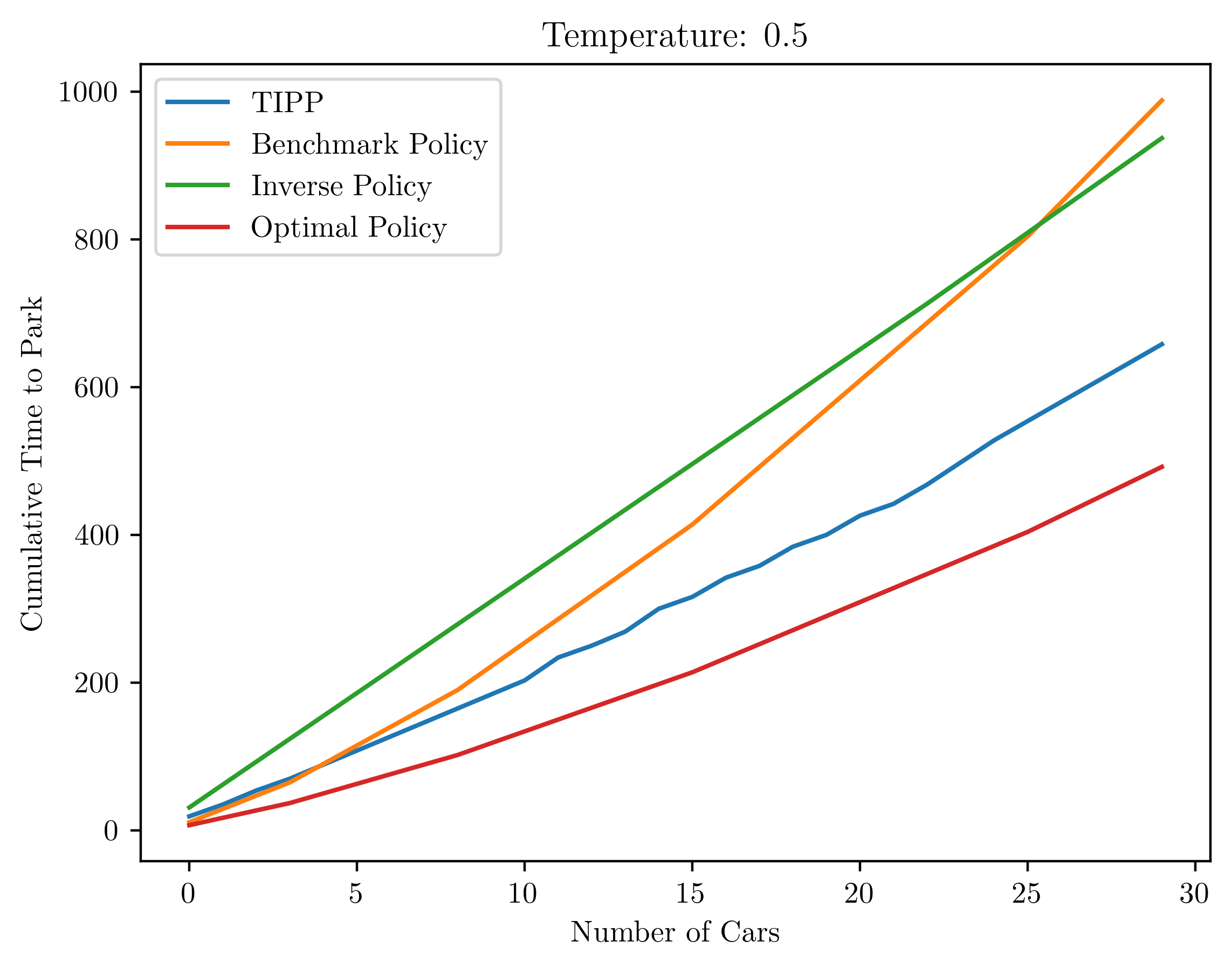}
  \label{fig:subfig2}
}
\hfill  % This ensures that the figures spread out to fill the line
\subfloat[Parking garage under high occupancy]{%
  \includegraphics[width=0.32\textwidth]{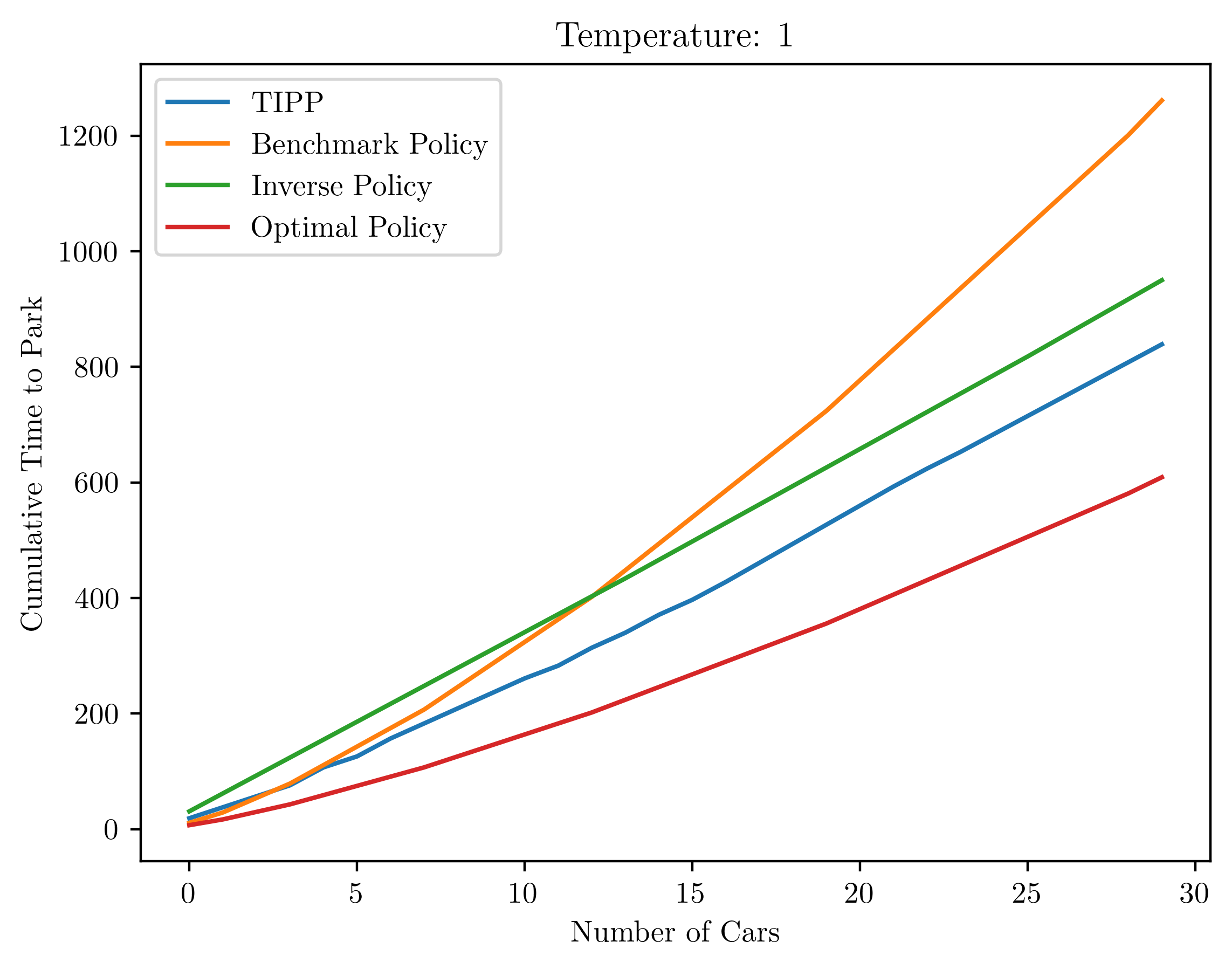}
  \label{fig:subfig3}
}
\caption{(a) Demonstrates the cumulative parking times in a garage under low occupancy (Temperature: 0.1), where the Benchmark policy initially outperforms TIPP. (b) and (c) Illustrate the effectiveness of the TIPP policy at medium (Temperature: 0.5) and high occupancy (Temperature: 1) levels respectively, indicating that TIPP excels when the occupancy level is sufficiently high to make parking decisions more challenging.}
\label{fig:cumulative}
\end{figure*}

To further evaluate the performance of TIPP under varying fill rates, we repeated the experiment across different temperatures, with the results displayed in Figure~\ref{fig:cumulative}. This figure presents the cumulative parking times, clearly demonstrating that TIPP outperforms both the Benchmark and Inverse policies. Notably, while the Benchmark Policy initially leads at a low temperature of 0.1, as shown in Figure~\ref{fig:cumulative_a}, TIPP surpasses it as the parking lot begins to fill. It's important to recognize that the effectiveness of a parking strategy becomes particularly significant not when the lot is empty and the empty spots are obvious, but rather when the parking situation becomes more congested and strategic decision-making is required.

\section{Conclusion}
% Yan: I have just write a draft of the conclusion part. Please feel free to revise it. 
In this study, we proposed an innovative approach to optimize vehicle parking efficiency using entropy-based dynamic programming. Our methodology addresses the challenges of parking congestion in urban environments by predicting the distribution of available parking spots across different levels of a multi-story parking garage using entropy models derived from statistical mechanics. 
We formulated the parking optimization problem as a dynamic program, aiming to minimize the time spent by drivers in finding parking spaces. Our approach considers factors such as driving time, walking time, and the probability of finding a parking spot on each floor.

Results from comparative analysis with other parking policies demonstrated the superiority of TIPP in terms of parking time efficiency. By dynamically guiding vehicles to the optimal floor, we significantly reduced the overall time spent searching for parking spots.

Future research can extend the methodology developed in this study in several promising directions. One possibility is to explore a more complex parking structure where individual parking spots are interconnected in a network, allowing the problem to be formulated as a Markov Decision Process (MDP). In such a scenario, an extension of the proposed entropy model can also be envisioned where every point of interest (i.e., storefront, park, attraction, etc.) contributes an additive energy term to each parking spot. Additionally, introducing stochasticity into the output of the model may enhance the balance between exploration and exploitation, potentially leading to more robust solutions under varying occupancy conditions.

\newpage

\bibliography{ref}

% Generated by IEEEtran.bst, version: 1.14 (2015/08/26)
\begin{thebibliography}{1}
\providecommand{\url}[1]{#1}
\csname url@samestyle\endcsname
\providecommand{\newblock}{\relax}
\providecommand{\bibinfo}[2]{#2}
\providecommand{\BIBentrySTDinterwordspacing}{\spaceskip=0pt\relax}
\providecommand{\BIBentryALTinterwordstretchfactor}{4}
\providecommand{\BIBentryALTinterwordspacing}{\spaceskip=\fontdimen2\font plus
\BIBentryALTinterwordstretchfactor\fontdimen3\font minus \fontdimen4\font\relax}
\providecommand{\BIBforeignlanguage}[2]{{%
\expandafter\ifx\csname l@#1\endcsname\relax
\typeout{** WARNING: IEEEtran.bst: No hyphenation pattern has been}%
\typeout{** loaded for the language `#1'. Using the pattern for}%
\typeout{** the default language instead.}%
\else
\language=\csname l@#1\endcsname
\fi
#2}}
\providecommand{\BIBdecl}{\relax}
\BIBdecl

\bibitem{inrix}
G.~Cookson and B.~Pishue, ``The impact of parking pain in the us, uk and germany car,'' Tech. Rep., 2017.

\bibitem{parking_games}
\BIBentryALTinterwordspacing
D.~Ayala, O.~Wolfson, B.~Xu, B.~Dasgupta, and J.~Lin, ``Parking slot assignment games,'' in \emph{Proceedings of the 19th ACM SIGSPATIAL International Conference on Advances in Geographic Information Systems}, ser. GIS '11.\hskip 1em plus 0.5em minus 0.4em\relax New York, NY, USA: Association for Computing Machinery, 2011, p. 299–308. [Online]. Available: \url{https://doi.org/10.1145/2093973.2094014}
\BIBentrySTDinterwordspacing

\bibitem{abidi2015}
\BIBentryALTinterwordspacing
S.~Abidi, S.~Krichen, E.~Alba, and J.~M. Molina, ``A new heuristic for solving the parking assignment problem,'' \emph{Procedia Computer Science}, vol.~60, pp. 312--321, 2015, knowledge-Based and Intelligent Information and Engineering Systems 19th Annual Conference, KES-2015, Singapore, September 2015 Proceedings. [Online]. Available: \url{https://www.sciencedirect.com/science/article/pii/S1877050915022590}
\BIBentrySTDinterwordspacing

\bibitem{chen2013}
I.~Chen~Zhirong, Xia Jianhong~(Cecilia) and Buntoro, ``Development of fuzzy logic forecast models for location-based parking finding services,'' \emph{Mathematical Problems in Engineering}, 2013.

\bibitem{yan2017}
\BIBentryALTinterwordspacing
Y.~Han, J.~Shan, M.~Wang, and G.~Yang, ``Optimization design and evaluation of parking route based on automatic assignment mechanism of parking lot,'' \emph{Advances in Mechanical Engineering}, vol.~9, no.~7, p. 1687814017712416, 2017. [Online]. Available: \url{https://doi.org/10.1177/1687814017712416}
\BIBentrySTDinterwordspacing

\bibitem{Zhang2022}
F.~L. Xinyuan~Zhang, Cong~Zhao, X.~Li, and Y.~Du, ``{Online parking assignment in an environment of partially connected vehicles: A multi-agent deep reinforcement learning approach},'' \emph{Transportation Research Part C}, 2022.

\bibitem{Khalid2022}
L.~W. Muhammad~Khalid, K.~Wang, N.~Aslam, C.~Pan, and Y.~Cao, ``Deep reinforcement learning-based long-range autonomous valet parking for smart cities,'' \emph{Sustainable Cities and Society}, 2022.

\bibitem{hu2019research}
X.~Hu, R.~Niu, and T.~Tao, ``Research on entropy based corrective maintenance difficulty estimation of metro signaling,'' in \emph{2019 IEEE Intelligent Transportation Systems Conference (ITSC)}.\hskip 1em plus 0.5em minus 0.4em\relax IEEE, 2019, pp. 79--85.

\bibitem{statmech}
K.~Huang, \emph{Statistical Mechanics}.\hskip 1em plus 0.5em minus 0.4em\relax Wiley, 1987.

\end{thebibliography}

% \begin{IEEEbiography}
% [{\includegraphics[width=1in,height=1.25in,clip,keepaspectratio]{Bio_pic/Yuhan.jpg}}]{Yuhan Tang} is a Master of Science student in Systems Engineering at the University of California, Berkeley.
% \end{IEEEbiography}

% that's all folks
\end{document}